\documentclass[12pt]{JHEP}
\usepackage{amssymb}
\def\Bbb{\mathbb}
\def\BZ{\mbox{$\Bbb Z$}} \def\BR{\mbox{$\Bbb R$}}
\def\BC{\mbox{$\Bbb C$}} \def\BP{\mbox{$\Bbb P$}}

\preprint{{\tt hep-th/0003242}\\ {\tt IMSc/2000/03/10} \\
March 27,2000}
\title{On the Landau-Ginzburg description of 
Boundary CFTs and special Lagrangian submanifolds}
\author{ Suresh Govindarajan\\
 Department of Physics\\
 Indian Institute of Technology, Madras \\
 Chennai 600 036 INDIA \\
 Email: \email{suresh@chaos.iitm.ernet.in}}
\author{T. Jayaraman \\
The Institute of Mathematical Sciences \\
C.I.T. Campus, Taramani \\
Chennai 600 113  INDIA \\ Email: \email{jayaram@imsc.ernet.in}}
\abstract{
We consider Landau-Ginzburg (LG) models with boundary conditions
preserving A-type $N=2$ supersymmetry. We show the equivalence of
a linear class of boundary conditions in the LG model to a
particular class of boundary states in the corresponding CFT by
an explicit computation of the open-string Witten index in the LG
model. We extend the linear class of boundary conditions to
general non-linear boundary conditions and determine their
consistency with A-type $N=2$ supersymmetry. This enables us to
provide a microscopic description of special Lagrangian
submanifolds in $\BC^n$ due to Harvey and Lawson. We generalise
this construction to the case of hypersurfaces in $\BP^n$. We
find that the boundary conditions must necessarily have vanishing
Poisson bracket with the combination
$(W(\phi)-\overline{W}(\overline{\phi}))$, where $W(\phi)$ is the
appropriate superpotential for the hypersurface.  An interesting
application considered is the $T^3$ supersymmetric cycle of the
quintic in the large complex structure limit. 
}

\begin{document}

\section{Introduction}

A complete microscopic description of D-branes wrapped on
supersymmetric cycles is available in the the cases where these
cycles are submanifolds in flat spaces like tori. The description
can also be fairly reliably extended to spaces where the
techniques of conformal field theories constructed from purely
free fields can be easily applied, as in the case of orbifolds.
However it is only recently that the case of D-branes living in
non-trivial curved spaces and wrapped on supersymmetric cycles in
these spaces have begun to be investigated systematically from a
microscopic viewpoint.  Following Ooguri et. al.\cite{ooy}, who
specified the boundary conditions on the worldsheet $N=2$
supersymmetry generators and explained their geometric
significance, further efforts have concentrated on extending the
boundary conformal field theory description of D-branes to the
case of Calabi-Yau (CY) 
manifolds\cite{RS,gutsat,quintic,oldpaper,gepnermodels,boundarycft}. 
Calabi-Yau manifolds in three complex dimensions have been the
subject of special attention in view of their importance of these
manifolds for string compactification. (See ref. \cite{strings}
for a nice summary. For earlier work that dealt with similar
issues without however explicitly describing D-branes, see ref.
\cite{sagnotti}.)

In the closed string case, string propagation on Calabi-Yau
manifolds can be described by a variety of techniques depending
on which region of the space of complex structure and K\"ahler
moduli of the CY manifold one wishes to concentrate on. At the
so-called Gepner point in the moduli space of some CY manifolds,
explicit descriptions are available in terms of the tensor
product of $N=2$ conformal field theories. This point can also be
described by using the Landau-Ginzburg (LG) description of these
conformal field theories. The LG description provides a link
between the abstract geometrical structure encoded in the CFT and
a more explicit description in terms of the co-ordinates of the
algebraic geometric picture of the CY manifold. The LG
description can be used also for CY manifolds that may not have a
corresponding Gepner construction. More generally, the LG models
may be viewed as the description appropriate to a particular
region in the enlarged moduli space of Calabi-Yau vacua.

For the study of D-branes one can use the corresponding
extensions of these descriptions to world-sheets with boundary.
In the case of the Gepner construction, one may use the boundary
conformal field theory techniques due to Cardy\cite{cardy}, to
provide an explicit construction of boundary states associated to
D-branes. However to make the geometric picture of D-branes more
explicit, one may, in simple cases, work with a functional
integral description of such theories with an explicit Lagrangian
involving free bosons and free fermions. For more complicated
examples of CY manifolds one would like to extend the LG
description to world-sheets with boundary. 
  
  Substantial progress has been achieved in the application of
the methods of boundary conformal field theory to the case of
D-branes wrapped on supersymmetric cycles in the CY. Following on
the work of Recknagel and Schomerus \cite{RS}that used the Gepner
model construction for the description of the boundary states
relevant to D-brane constructions on supersymmetric cycles on
Calabi-Yau manifolds, the specific case of D-branes on the
quintic Calabi-Yau manifold was studied in detail in the work of
Brunner et.al.\cite{quintic}. Among other results, a particularly
important one (and relevant to the results of this paper as we
shall explain below) was their use of the identification of the
Witten index in the open string sector between two boundaries
with the intersection matrix between the corresponding D-branes
to study systematically some properties of D-branes with both
A-type and B-type boundary conditions (in the notation of Ooguri
et.al \cite{ooy}). Subsequent papers have utilised these
techniques to particularly study B-type boundary states in other
Calabi-Yau manifolds\cite{gepnermodels}.

  Despite this impressive progress, several important puzzles and
open questions remain. It would take us too far afield to list
these but there are two that are the underlying theme of the
present paper. What precisely is the geometric interpretation of
the large number of D-brane like boundary states that are to be
found in the boundary CFTs arising from Gepner type
constructions? Secondly, are there more general geometric
constructions that may or may not be realised in the boundary
conformal field theory approach? While the answer to the second
question is generally yes, we still need to explicitly
investigate such constructions. We note that these questions need
to be clarified further separately in the case of A-type and
B-type boundary conditions. For A-type boundary conditions one
may appeal to the {\it modified geometric hypothesis} of ref.
\cite{quintic,strings}. According to this hypothesis we may
expect that the masses and charges for branes with A-type
boundary conditions conditions computed in the ``large volume''
limit continue to hold in the ``small volume'' limit also. Thus
the A-type boundary conditions can in principle be computed in a
suitable description that keeps explicit track of the geometry
associated with the corresponding Gepner construction (modulo
some caveats that we shall discuss later). In the case of B-type
branes this is not expected to be true. Following the method
developed by Brunner et. al. the data at the Gepner point have to
be monodromy transformed to the large volume limit (by using the
monodromy transformations computed in the mirror CY) to obtain
the corresponding interpretation of these branes. In the case of
B-type boundary conditions the corresponding charges in the large
volume descriptions have been obtained of all the boundary states
obtained in several examples including the quintic.  However a
full geometric or physical understanding is still lacking,
particularly with regard to the description at the Gepner point.
In the case of the A-type boundary conditions only one class of
boundary states have been tentatively identified with the
corresponding geometric construction. 

  In this paper, as a first step in trying to answer these
questions, we will investigate in detail general classes of
A-type boundary conditions from a more geometric viewpoint. This
will lead us to not only investigate boundary conditions related
to the Recknagel-Schomerus construction but also more general
constructions that clearly go beyond the Recknagel-Schomerus
class. 
 
  In an earlier paper\cite{oldpaper}, the correspondence between
boundary states in boundary CFT and boundary conditions in LG
models was studied. This correspondence was explicitly
illustrated in the case of the supersymmetric one-cycles of the
two-torus, using the common discrete symmetries of the boundary
conformal field theory and the boundary LG theory. A general
class of linear boundary conditions in the LG models was also
described.  These are relevant to both the case of D-branes
wrapped on the middle-dimensional cycles of a CY as well as the
case of even-dimensional D-branes wrapped on holomorphic
sub-manifolds of a CY. However an explicit identification of the
boundary states of the CFT with those from the LG theory was
still lacking.

   In this paper we will begin by considering linear A-type
boundary conditions in LG models, of the form discussed in our
earlier paper. By explicitly performing the open-string Witten
index calculation in the LG model and comparing it to the
boundary conformal field theory calculation we will definitively
show the equivalence of this class of linear boundary conditions
with the $L=\lfloor k/2\rfloor$ class of boundary states in the
boundary CFT.

We will then turn to more general, generically non-linear, A-type
boundary conditions and show the consistency conditions that are
required to ensure that these describe supersymmetric
middle-dimensional cycles in Calabi-Yau manifolds. It is well
known that in the case of $N=2$ world-sheet supersymmetry several
interesting features of the conformal limit are seen even when
the theory is perturbed away from this limit. Hence models with
$N=2$ world-sheet supersymmetry even away from the conformal
limit are of interest. In our discussion therefore we will
consider A-type boundary conditions in Landau-Ginzburg
descriptions of minimal models, both without and with
perturbations by relevant operators. We extend this discussion to
more general cases.

A summary of the main results of the paper is as follows:
\begin{enumerate}

\item We compute the open-string Witten index in the LG model and
provide evidence that the linear class of boundary conditions in
the minimal model correspond to the $L=\lfloor k/2\rfloor$
boundary states in the minimal model.
\item We show that A-type $N=2$ supersymmetry is preserved if the
submanifold is {\it Lagrangian}. The complete set of boundary
conditions associated with this Lagrangian submanifold are
presented. The analysis is used to provide a microscopic
description of the special Lagrangian submanifolds in $\BC^n$ due
to Harvey and Lawson\cite{harveylawson}. 
\item For the cases with a superpotential $W$ (which describe
hypersurfaces in $\BP^n$), we show that one needs to have the
boundary conditions have vanishing Poisson bracket with
$(W-\overline{W})$. Thus, these submanifolds are necessarily
pre-images of {\it straight lines} in the $W$-plane.
\item For a single minimal model, we find non-linear boundary
conditions by perturbations of the quasi-homogeneous potential by
relevant operators. The boundary conditions correspond to
straight lines in the $W$-plane passing through the minima of the
perturbed potential. 
\item For the case of the quintic CY threefold, we use these
methods to provide an explicit microscopic description of the
$T^3$ special Lagrangian sub-manifold in the infinite complex
structure limit. 
\end{enumerate}
Observations closely related to items 3 and 4 above  from a slightly
different viewpoint\footnote{These results have been reported by C. Vafa
in a recent conference talk\cite{vafatalk}. We thank C. Vafa for
bringing this to our attention.}  appear in the work of Hori, Iqbal and
Vafa\cite{hiv}.  Some of these results have been reported in
a recent paper by Hori and Vafa which
appeared while this manuscript was under
preparation\cite{horivafa}.) 

  The organisation of the paper is as follows: In section 2, we
discuss the case of LG models with boundary and list a linear
class of boundary conditions obtained in \cite{oldpaper}. In
section 3, we review the construction of A-type boundary states
in a single minimal model using Cardy's prescription. In section
4, we carry out the open-string Witten index computation and
provide a map from the boundary conditions in the LG model to a
class of boundary states in the corresponding CFT. This is
illustrated for the case of a single minimal model and for the
Gepner model associated with the quintic. In section 5, we
consider general boundary conditions consistent with A-type $N=2$
worldsheet supersymmetry. Using this microscopic description, we
obtain conditions under which the boundary conditions describe a
supersymmetric cycle (special Lagrangian).  We apply the methods
to some simple examples. We conclude in section 6 with some
remarks.

\section{Landau-Ginzburg theories with boundary}

\subsection{Notation and Conventions}

We work in $N=2$ superspace with coordinates $x^m$,
$\theta^\alpha$, $\overline{\theta}^{\dot\alpha}$ ($m=0,1$,
$\alpha,{\dot\alpha}=+,-)$. Left movers are specified by the
index $-$ and right movers by the index $+$. The worldsheet has
Lorentzian signature (metric=Diag$(-1,+1)$) and has a boundary at
$x^1=0$ and is topologically a half-plane. 

The Lagrangian for a $N=2$ supersymmetric Landau-Ginzburg theory
is constructed from chiral superfields 
($y^m=x^m+i\theta^\alpha \sigma^m_{\alpha\dot\alpha}
\overline{\theta}^{\dot\alpha}$) 
$$\Phi(x,\theta) = \phi(y) + \sqrt2
\theta^\alpha \psi_\alpha(y) + \theta^\alpha \theta_\alpha F(y)$$
and anti-chiral superfields. The Lagrangian for $n$ chiral
superfields $\Phi_i$ is given by
\begin{equation}
S=\int d^2x d^4\theta K(\Phi,\overline{\Phi}) -
\int d^2x d\theta^+d\theta^- W(\Phi) -
\int d^2x  d\overline{\theta}^+d\overline{\theta}^- \overline{W}(\overline{\Phi})
\quad,
\end{equation}
where $K$ is the K\"ahler potential and $W$ is the holomorphic
superpotential. We will choose the K\"ahler potential to be
$K=\sum_i \overline{\Phi}_i\Phi_i$. In the conformal case, the
superpotential is taken to be quasi-homogeneous: $W(\lambda^{n_i}
\Phi_i) =\lambda^d W(\Phi_i)$, where $n_i$ are some integers
which are related to the charges of the superfields $\Phi_i$. The
Lagrangian takes the following form after the auxiliary fields
$F_i$ are eliminated\footnote{In addition, we have symmetrised
the action of the derivatives occuring in the Kinetic energy term
for the fermions as is done when one is considering worldsheets
with boundary.}
\begin{eqnarray}
S = \int d^2x \left( -\partial_m\overline{\phi}_i
\partial^m \phi_i + 
i\overline{\psi}_{-i}(\stackrel{\leftrightarrow}{\partial_0} + 
\stackrel{\leftrightarrow}{\partial_1})\psi_{-i}
+ i\overline{\psi}_{+i}(
\stackrel{\leftrightarrow}{\partial_0} -
\stackrel{\leftrightarrow}{\partial_1})\psi_{+i} \nonumber
\phantom{{\partial A}\over {\partial B}}\right.\\
-\left. \left|{{\partial W}\over {\partial\phi_i}}\right|^2 
-{{\partial^2 W}\over{\partial\phi_i\partial\phi_j}} \psi_{-i}\psi_{+j}
-{{\partial^2
\overline{W}}\over{\partial\overline{\phi}_i\partial\overline{\phi}_j}} 
\overline{\psi}_{+i}\overline{\psi}_{-j}\right)\quad,
\end{eqnarray}
where $A \stackrel{\leftrightarrow}{\partial_i} B \equiv 
{1\over2}[A (\partial_i B) - (\partial_i A) B]$.

The Lagrangian is invariant under the supersymmetry
transformations parametrised by $\epsilon_\alpha$ and
$\overline{\epsilon}_{\dot{\alpha}}$.  The transformations of the
fields are given by
\begin{eqnarray}
\delta\phi_i &=&\sqrt2 (-\epsilon_- \psi_{+i} + \epsilon_+\psi_{-i})
\nonumber \\
\delta\psi_{+i} &=& i\sqrt2 (\partial_0 + \partial_1)\phi_i
\overline{\epsilon}_- + \sqrt2
\epsilon_+ {{\partial \overline{W}}\over {\partial\overline{\phi}_i}} 
\label{susy} \\
\delta\psi_{-i} &=& -i\sqrt2 (\partial_0 - \partial_1)\phi_i 
\overline{\epsilon}_++ \sqrt2
\epsilon_- {{\partial \overline{W}}\over {\partial\overline{\phi}_i}} 
\nonumber
\end{eqnarray}

We will be interested in considering the case when boundary
conditions preserve part of the supersymmetry. Further, the
boundary conditions should cancel the ordinary variations of the
action modulo the bulk equations of motion. The ordinary
variation of the action gives rise to the following boundary
terms
\begin{eqnarray}
\delta_{\rm ord} S = \int dx^0 &&\left ( [(\partial_1
\overline{\phi}_i) \delta \phi + \delta \overline{\phi}_i (\partial_1
\phi_i)]|_{x^1=0} \phantom{A\over B}\right. \nonumber \\ 
 + &&\left. {i\over2} [
\delta \overline{\psi}_{-i} \psi_{-i}
-\overline{\psi}_{-i}\delta  \psi_{-i}
-\delta \overline{\psi}_{+i} \psi_{+i}
+\overline{\psi}_{+i}\delta  \psi_{+i}]|_{x^1=0} \right)
\label{ord}
\end{eqnarray}

There are two inequivalent sets of boundary conditions which
preserve different linear combinations of the left and right
$N=2$ supersymmetries\cite{ooy,warner}. 

\noindent {\bf A-type boundary conditions:} These are boundary
conditions such that the unbroken $N=2$ supersymmetry is
generated by
\begin{equation}
\epsilon_+ = \eta\ \overline{\epsilon}_- \quad,
\end{equation}
and the complex conjugate equation and $\eta=\pm1$ corresponds to
the choice of spin-structure on the worldsheet. 

\noindent{\bf B-type boundary conditions:} 
These are boundary conditions such that the unbroken $N=2$
supersymmetry is generated by
\begin{equation}
\epsilon_+ = \eta\ \epsilon_- \quad,
\end{equation}
and the complex conjugate equation and $\eta=\pm1$ corresponds to
the choice of spin-structure on the worldsheet. The two boundary
conditions are related by the mirror automorphism of the $N=2$
supersymmetry algebra under which the left-moving $U(1)$ current
changes sign. 

\subsection{A-type boundary conditions}

Under A-type boundary conditions, the unbroken $N=2$ supersymmetry
is given by the condition
\begin{equation}
\epsilon_+ = \eta\ \overline{\epsilon}_-\quad,
\end{equation}
where $\eta=\pm1$.  In an earlier paper, it was shown that the following
conditions\footnote{The boundary conditions have been adapted to the
notation used in this paper.}
preserve $N=2$ supersymmetry and that the boundary terms in
ordinary variations (eqn. (\ref{ord})) of the Lagrangian vanish.
\begin{eqnarray}
(\psi_{+i} - A_{ij} \eta\ \overline{\psi}_{-j})|_{x^1=0} =0\quad, \nonumber \\
\partial_1 (\phi_i + A_{ij} \overline{\phi}_j)|_{x^1=0} =0 \quad,\nonumber \\
\partial_0 (\phi_i - A_{ij} \overline{\phi}_j)|_{x^1=0} =0 \quad,\\
\left.\left(A_{ij} {{\partial W}\over{\partial\phi_j}} -
 {{\partial \overline{W}}\over{\partial\overline{\phi}_i}}
\right)\right|_{x^1=0} =0
\nonumber\quad,
\end{eqnarray}
where $A$ is a symmetric matrix satisfying $AA^{\dagger}=1$.

For the $k$-th minimal model, the LG description
has a superpotential given by $W=\phi^{k+2}/(k+2)$, the condition involving
the superpotential becomes
\begin{equation}
A^{k+2} =1\quad,
\end{equation}
which is a condition on the parameter $A$ appearing in the
boundary condition. Thus, $A$ can be any $(k+2)$-th root of
unity.  Hence there are $(k+2)$ different boundary conditions
which are consistent with $N=2$ supersymmetry. 

Under the action of the generator $g$ of the group $Z_{k+2}$, one
can easily check that $A\rightarrow A \exp(4\pi i/(k+2))$.
Suppose we choose to label the different $(k+2)$ roots of unity
by
$$A_m = \exp (2\pi m/k+2)\quad.$$ 
Then under the action of $g$, $A_m \rightarrow A_{m+2}$.  This
suggests that the $m$ label here can be associated with the $M$
labels of the boundary states constructed in the corresponding
minimal model. For odd $k$, the allowed values of $A$ form a
$(k+2)$ dimensional orbit while for even $k=2n$, one obtains two
$(n+1)$ dimensional orbits of the $Z_{n+1}$ subgroup of the
$Z_{k+2}$. 

\subsection{B-type boundary conditions}

Under B-type boundary conditions, the unbroken $N=2$ supersymmetry
is given by the condition
\begin{equation}
\epsilon_+ = \eta\ \epsilon_-\quad,
\end{equation}
where $\eta=\pm1$. The following linear boundary conditions
were constructed in the LG model\cite{oldpaper} 
\begin{eqnarray}
(\psi_{+i} + \eta {B_i}^j\psi_{-j})|_{x=0}=0\quad, \nonumber  \\
\partial_1(\phi_i + {B_i}^j\phi_j)|_{x=0}=0\quad, \nonumber  \\
 \partial_0(\phi_i - {B_i}^j\phi_j)|_{x=0}=0\quad, \nonumber  \\
\left.\left({{\partial W}\over{\partial\phi_i}} + {B^*_i}^j {{\partial
W}\over{\partial\phi_j}}\right)\right|_{x=0}=0 \quad,
\end{eqnarray}
where the boundary condition is specified by a hermitian matrix
$B$ which satisfies $B^2=1$.  Since $B$ squares to one, its
eigenvalues are $\pm1$.  An eigenvector of $B$ with eigenvalue of
$+1$ corresponds to a Neumann boundary condition and $-1$
corresponds to a Dirichlet boundary condition. Associated with
every eigenvector with eigenvalue $+1$, there is a non-trivial
condition involving the superpotential which is given by the last
of the above boundary conditions. 

For a LG model with a single chiral superfield such as the
minimal model, the consistency condition involving the
superpotential does not permit the imposition of a Neumann
boundary condition on the scalar field. Thus, one can only impose
Dirichlet boundary conditions on the scalar. We will not consider
B-type boundary states further in this paper.

\section{Boundary States in the $N=2$ minimal models}

\subsection{Notation and Conventions}

The $k$-th $N=2$ minimal models  has central charge
$c=3k/(k+2)$. The primary fields of the model are labelled by
three integers $(l,m,s)$ with 
\begin{eqnarray*}
l&=&0,\cdots, k\quad,  \\
m&=&-(k+1),-k,\cdots,(k+2)\ {\rm mod}\ (2k+4)\quad, \\
 s&=&-1,0,1,2\ {\rm mod}\ 4\quad, 
\end{eqnarray*}
subject to the constraint that $l+m+s$ is even. In addition there
is a field identification given by $$(l,m,s)\sim
(k-l,m+k+2,s+2)\quad.$$ Even $s$ refers to the NS sector and odd
$s$ refers to the R sector fields. 

A complete set of labels for the minimal model (using the field
identification mentioned above) are given by
$$
l=0,\cdots, \lfloor k/2\rfloor\quad,\quad
m=-(k+1),-k,\cdots,(k+2)\quad {\rm and} \quad s=-1,0,1,2\quad,
$$
where $\lfloor k/2\rfloor$ is the largest integer less than or
equal to $k/2$ and $(l+m+s)$ is even.

Another equivalent set of labels is given by
$$
l=0,\cdots, k\quad,\quad
m=-(k+1),-k,\cdots,(k+2)\quad {\rm and} \quad s=0,1\quad,$$
where again we have the condition that $(l+m+s)$ must be even.
The dimension $h$ and $U(1)$ charge $q$ of the fields are given by
\begin{eqnarray}
h_{l,m,s}&=&\frac{l(l+2) - m^2}{4(k+2)}~+~\frac{s^2}{8}
\nonumber \\
q_{l,m,s}&=&\frac{m}{k+2} - \frac{s}{2}\quad {\rm mod}~2
\end{eqnarray}

The $k$-th minimal model has a $\BZ_{k+2}\times \BZ_2$ discrete
symmetry. The action of the discrete symmetry on the fields is given by
\begin{eqnarray}
g\cdot \Phi_{l,m,s} = e^{{2\pi i m}\over{k+2}}\ \Phi_{l,m,s}\quad,
\\
h \cdot \Phi_{l,m,s} = (-)^s\ \Phi_{l,m,s}\quad,
\end{eqnarray}
where $g$ and $h$ generate the $\BZ_{k+2}$ and $\BZ_2$ respectively.
We will be interested in the action of $\exp(i\pi J_0)$ on a bulk state:
\begin{equation}
\exp(i\pi J_0)\ |l,m,s\rangle = \exp(i\pi [{m\over{k+2}} -{s\over2}])
\ |l,m,s\rangle \quad. 
\end{equation}
Note that this is not necessarily equal to $(-)^{F_L}$ when one
is considering a single minimal model (as opposed to a Gepner
construction involving integer $U(1)$ charges). However, one can
see that product $(-)^{F_L}\exp(i\pi J_0)$ commutes with all
generators of the $N=2$ supersymmetry algebra. In the minimal
model, this product on general grounds should be given by
$f(g,h)$, where $f$ is some function of the the discrete
symmetries which commute with $N=2$ supersymmetry\cite{wittenlg}.
Thus, we will also be interested in defining the action of
$(-)^{F_L}$ on the states $|l,m,s\rangle$. We will require that
$(-)^{F_L}$ gives $\pm1$ acting on the NS sector states. For odd
$k$, we find
\begin{equation}
(-)^{F_L}|l,m,s\rangle  = \exp(i\pi [m+{s\over2}])\ |l,m,s\rangle \quad. 
\end{equation}
The above assignment is consistent with the identification
$|l,m,s\rangle\sim |k-l,m+k+2,s+2\rangle$ of states. Thus, for
odd $k$, $f(g,h)= g^{(k+3)/2}$. 

For a given representation $p$ of the $N=2$ algebra, the character is
defined as
\begin{equation}
\chi_p\left(q, z, u \right)~=~e^{-2 i \pi u}\ {\rm Tr}_p\ [e^{2 i \pi z 
J_0}\ q^{(L_0-{c \over 24})}]
\end{equation}
where $q=\exp(2i\pi\tau)$ and $u$ is an arbitrary phase.  The
trace runs over the representation denoted by $p$. The characters
of the $N=2$ minimal models are defined in terms of the Jacobi
theta functions $\theta_{n,m}(\tau,z,u)$ and characters of a
related parafermionic theory $C^l_m(\tau)$ as: 
\begin{equation}
\chi_{l,m}^{(s)}\left(q, z, u \right) = \sum_{j~{\rm mod}~k}
C^l_{m+4j-s} (\tau)\ \theta_{2m+(4j-s)(k+2),2k(k+2)}\left(\tau, 2kz, u
\right)\quad.
\end{equation}
The characters $\chi_{l,m}^{(s)}$ have the property that they are invariant
under $s \rightarrow s+4$ and $m \rightarrow m+2(k+2)$ and are zero 
if $(l+m+s)$ is odd. By using the properties of the theta functions, 
the modular transformation of the minimal model characters is found to be
\begin{equation}
\chi_{l,m}^{(s)} \left( \hat{q},0,0 \right) = C
\sum_{l',m',s'} \sin(l,l')_k \exp \left({i \pi mm' \over k+2}\right)
\exp\left(- {i \pi ss' \over 2} \right) \chi_{l',m'}^{(s')}(q,0,0)
\end{equation}
where $\hat{q}=\exp(-2i\pi/\tau)$;
$(l,l')_k \equiv \left({\pi(l+1)(l'+1) \over k+2} \right)$
and $C=1/\sqrt2(k+2)$.

\subsection{A-type Boundary States in the $N=2$ minimal models}

We will consider A-series which has a diagonal partition
function.  For A-type boundary conditions, there are Ishibashi
states\cite{ishibashi} for each possible value of $(l,m,s)$. We
will label these states by $|l,m,s\rangle\rangle$. Using Cardy's
prescription\cite{cardy}, we can construct the boundary states
\begin{equation}
|L,M,S\rangle = \sqrt{C}\sum_{l,m,s} 
{{\sin(L,l)_k} \over
{~~\left[\sin(l,0)_k\right]^{- {1\over2}} }}
\exp\left({{i\pi Mm}\over{k+2}}\right) 
\exp\left(- {i \pi Ss \over 2}\right) |l,m,s\rangle\rangle
\end{equation}
where we have used upper case letters to represent the boundary
state and lower case letters for the Ishibashi states.  One can
check that the boundary states $|L,M,S\rangle$ and
$|L,M,S+2\rangle$ differ only in the sign occurring in front of
the RR-sector (i.e., odd $s$ ) Ishibashi states. Thus, it
suffices to study only the $S=0,1$ states.

The field identification $(l,m,s)\sim (k-l,m+k+2,s+2)$ in the
bulk minimal model extends to the boundary states as
$|L,M,S\rangle\sim |k-L,M+k+2,S+2\rangle$.  The annulus amplitude
${\cal A}_{L,M,S}(q)$ (with modular parameter $q$)  which is
given by the modular transform of the cylinder amplitude $
\langle 0,0,0|\hat{q}^{L_0 - c/24}|L,M,S\rangle $, is given by
\begin{equation}
{\cal A}_{L,M,S}(q) = \chi_{L,M}^{(S)}(q)\quad.
\end{equation}
Note that this vanishes when $(L+M+S)$ is odd. Thus, we impose
the additional condition that $(L+M+S)$ be even.

The full set of boundary states that we obtain are specified by
the following values of $(L,M,S)$: 
$$
L=0,\cdots, \lfloor k/2\rfloor\quad,\quad
M=-(k+1),-k,\cdots,(k+2)\quad {\rm and} \quad S=0,2\quad,
$$
In this labelling convention, we will sometimes loosely refer to
the $S=2$ state as an antibrane (in analogy with the situation in
the full Gepner construction) since the $S=2$ boundary state
differs from the $S=0$ state (for identical values of $L,M$)  by
an overall sign in front of the RR Ishibashi states. This set of
labels takes care of the identification of boundary states
mentioned earlier except for the case when $k$ is even and
$L=k/2$. For this case, the antibrane corresponding to the
boundary state $|k/2,M,0\rangle$ is $|k/2,M+k+2,0\rangle$.

Under the discrete symmetries of the minimal model
$\BZ_{k+2}\times \BZ_2$, the boundary states transform as
\begin{eqnarray}
g\cdot |L,M,S\rangle = |L,M+2,S\rangle \\
h\cdot|L,M,S\rangle =|L,M,S+2\rangle
\end{eqnarray}
Thus all A-type boundary states can be classified into orbits of
the discrete symmetry. When $k$ is odd, there are $ \lfloor
k/2\rfloor =(k-1)/2$ orbits of length $(k+2)$ after taking into
account the identification of the boundary states mentioned
earlier. For even $k$, when $l<k/2$, then the states are in
orbits of length $(k+2)$. However, when $l=k/2$, since $l=k-l$,
the orbit length is shorter and equals $(k+2)/2$ (provided one
ignores the distinction between the $S=0$ and $S=2$ states). 

The characters of the full $N=2$ supersymmetry algebra are given
by the combination $(\chi_{lm}^{(s)} + \chi_{lm}^{(s+2)})$. It is
thus of interest to construct boundary states corresponding to
these characters. In this regard consider
\begin{equation}
|L,M,\pm\rangle \equiv {1\over\sqrt2}(|L,M,S\rangle \pm
|L,M,S+2\rangle)\quad.
\end{equation}
From the earlier discussion, it is clear that the states
$|L,M,+\rangle$ involve Ishibashi states from the NSNS sector and
$|L,M,-\rangle$ involve Ishibashi states from the RR sector.
These states also are more natural in the construction of
boundary states in the Gepner model since the tensor product of
boundary states $\prod_i |L_i,M_i,+\rangle$ incorporates the
condition that NSNS states of each sub-theory (labelled by $i$) 
are tensored to each other and the tensor product of boundary
states $\prod_i |L_i,M_i,-\rangle$ works similarly for the RR
states.  The annulus amplitude ${\cal A}_{L,M,\pm}(q)$ (with
modular parameter $q$)  which is given by the modular transform
of the cylinder amplitude $ \langle 0,0,+|\hat{q}^{L_0 -
c/24}|L,M,\pm\rangle $, is given by
\begin{equation}
{\cal A}_{L,M,\pm}(q) = \chi_{L,M}^{(S)}(q) \pm \chi_{L,M}^{(S+2)}(q) \quad,
\end{equation}
where $S=L+M$ mod $2$.  Under the discrete symmetries of the
minimal model $\BZ_{k+2}\times \BZ_2$, the boundary states
transform as
\begin{eqnarray}
g\cdot |L,M,\pm\rangle &=& |L,M+2,\pm\rangle \\
h\cdot|L,M,\pm\rangle &=&\pm |L,M,\pm\rangle
\end{eqnarray}
As before, all states except the case when $k$ is even and
$L=k/2$, the states can be arranged in $\BZ_{k+2}$ orbits.
However, when $k$ is even and $L=k/2=n$, the orbit length is
shorter. One has
$$
g^{n+1}\cdot |L,M,\pm\rangle = \pm |L,M,\pm\rangle\quad,
$$
Thus, they have orbit length $(n+1)=(k+2)/2$.

\section{Computing the open-string Witten index}

	We have so far constructed A-type boundary conditions in
the Landau-Ginzburg model and constructed A-type boundary states
in the corresponding minimal model. However, since the number of
boundary conditions is smaller than the number of states, we
would like to identify to which boundary states to which they
correspond.  This is not easy to do even in simple cases such as
the Ising model with boundary. A useful tool in this regard is to
classify boundary conditions and boundary states in terms of
discrete symmetries such as the $\BZ_2$ in the Ising model. In
the present problem, as we have already seen, there is a
$\BZ_{k+2}$ discrete group which organises the boundary
conditions and boundary states into orbits. It turns out that
this alone is sufficient to provide an identification for the
even $k$ minimal model. The LG boundary conditions form two
orbits of the $\BZ_{(k+2)/2}$ subgroup of $\BZ_{k+2}$. This
uniquely identifies them with the $L=k/2$ boundary states. 

This is not the case for odd $k$ where the LG boundary conditions
form a single $Z_{k+2}$ orbit which is true of all boundary
states in the corresponding minimal model. In order to make the
identification, we will use a open-string Witten index
computation (due to Douglas and Fiol\cite{douglasfiol}). In the
context of Calabi-Yau threefolds, this index computes the
intersection matrix between three cycles. We will compute the
index in both the LG as well as boundary CFT and show that the LG
boundary conditions correspond to the $L=\lfloor k/2\rfloor$
boundary states.

Let $|B\rangle$ and $|B'\rangle$ be two boundary states. The
Witten index is defined as\cite{douglasfiol}
\begin{equation}
\widetilde{\cal I}_{BB'} = {}_{RR}\langle B'|\ (-)^{F_L}  \
\tilde{q}^{(L_0 -c/24)}\ |B\rangle_{RR}\quad,
\end{equation}
where $|B\rangle_{RR}$ refers to the RR part of the boundary state.
In the open-string channel, this counts the number of {\it Ramond ground
states} of the Hamiltonian $H_{BB'}$:
\begin{equation}
\widetilde{\cal I}_{BB'} = {\rm Tr}_{BB'} \left[(-)^F q^{(L_0
-c/24)}\right]\quad.
\end{equation}
As discussed by Witten\cite{wittenlg}, the operator $(-)^{F_L}$
can be replaced by $\exp(i\pi J_0^L)$ where $J_0^L$ is the
zero-mode of the left-moving $U(1)$ current.  Thus, we will be
computing the following object
\begin{equation}
{\cal I}_{BB'} = {}_{RR}\langle B' |\  e^{i \pi J_0^L}\ \hat{q}^{(L_0
-c/24)}\  |B\rangle_{RR}\quad.
\end{equation}
In the open-string channel, this will be given by
\begin{equation}
{\cal I}_{BB'} = {\rm Tr}_{BB'} \left[e^{i \pi J_0 }
q^{(L_0 -c/24)}\right] \quad,
\end{equation}
where $J_0$ is the charge associated with the unbroken U(1).

For the level $k$ ($k$ odd) minimal model, one can study the
action of $(-)^{F_L}$ and $\exp(i \pi J_0^L)$ on the boundary
states. For A-type boundary states, we can see that
\begin{eqnarray}
\exp(i \pi J_0^L) |L,M,S\rangle &=& |L,M+1,S+1\rangle \\
(-)^{F_L} |L,M,S\rangle &=& |L,M+k+2,S-1\rangle\quad.
\end{eqnarray}
Thus, we see that for A-type boundary states (and odd $k$)
\begin{equation}
\exp(i \pi J_0^L) = g^{{k+3}\over2}h (-)^{F_L}\quad.
\label{relate}
\end{equation}

\subsection{Boundary Minimal Model Calculation}

We will now compute the following in the boundary minimal model
\begin{equation}
{\cal I}_{L,M,0;L',M',0} \equiv {}_{RR}\langle L',M',0|\ \exp(i \pi J_0)
\ \hat{q}^{(L_0 -c/24)}\  |
L,M,0\rangle_{RR}
\quad.
\end{equation}
This calculation is identical to the one in the appendix of ref. 
\cite{quintic} tailored to the case of a single minimal model. We
reproduce it here for completeness.  Using the expression for the
boundary states constructed using Cardy's prescription, we get
\begin{equation}
{\cal I}_{L,M,0;L',M',0} = 
 C  \sum_{l,m,s}^{R} {{\sin(L,l)_k \sin(L',l)_k}\over{\sin(l,0)_k}}
\exp\left({{i\pi(M-M'+1)m}\over{k+2}}\right) 
e^{-i\pi s/2}
\chi_{lm}^s(\hat{q})\quad,
\end{equation}
where the $R$ in the summation refers to the restriction to the
Ramond sector (i.e, $s=\pm1$).  On transforming to the
open-string channel by an S-transformation, one obtains
\begin{eqnarray}
{\cal I}_{L,M,0;L',M',0} &=& 
C^2 \sum_{l,m,s}^{R} \sum_{l',m',s'}
{{\sin(L,l)_k \sin(L',l)_k\sin(l.l')_k}\over{\sin(l,0)_k}}
e^{\left({{i\pi\mu m}\over{k+2}}\right)} 
e^{-i\pi s(1+s')/2}
\chi_{l'm'}^{s'}(q)\quad, \nonumber \\
&=& -2C^2 \sum_{l,m}^{R}\sum_{l',m'}^{R}
{{\sin(L,l)_k \sin(L',l)_k\sin(l.l')_k}\over{\sin(l,0)_k}}
e^{\left({{i\pi\mu m}\over{k+2}}\right)}   I_{l'}^{m'}(q)\quad,
\end{eqnarray}
where $\mu\equiv M-M'+m'+1$ and 
$$ 
I_l^m(q)\equiv
\chi_{l,m}^{(1)}(q)-\chi_{l,m}^{(-1)}(q)=\delta_{m,l+1}-\delta_{m,-l-1}\quad.
$$
(See ref. \cite{yang} for the above relation.)  In the above, we
have carried out the $s$ and $s'$ summations and hence the
restriction $R$ now implies that $(l+m)$ and $(l'+m')$ must be
odd.  On carrying out the summation over $m$, we get
\begin{equation}
{\cal I}_{L,M,0;L',M',0} = -2C^2 (k+2) \sum_{l=0}^k \sum_{l',m'}^{R}
{{\sin(L,l)_k \sin(L',l)_k\sin(l.l')_k}\over{\sin(l,0)_k}}
\delta_{\mu}^{(k+2)} (-)^{{\mu(l+1)}\over{(k+2)}} I_{l'}^{m'}(q)\quad,
\end{equation}
where $\delta_{\mu,0}^{(k+2)}$ is the periodic delta function of
period $(k+2)$ i.e, it is non-vanishing for $\mu=0$ mod $(k+2)$.
We can now carry out the summation over $l$. We then obtain
\begin{equation}
{\cal I}_{L,M,0;L',M',0}=-C^2 (k+2)^2 \sum_{l',m'}^{R} N_{LL'}^{l'}
\delta_{M-M'+m'+1}^{(2k+4)} I_{l'}^{m'}(q)\quad,
\end{equation}
where $N_{LL'}^{l'}$ is the $SU(2)$ level $k$ fusion coefficient
and $\delta_{M}^{(2k+4)}$ is the periodic delta function with
period $(2k+4)$. On carrying out the summations over $l'$ and
$m'$ we obtain (after substituting for $C$) 
\begin{equation}
{\cal I}_{L,M,0;L',M',0} = N_{L,L'}^{M-M'}\quad,
\end{equation}
where we have continued the top index $M$ of the $SU(2)$ fusion
coefficient $N_{L,L'}^M$ to values mod $(2k+4)$ following the
work of Brunner et. al.\cite{quintic}. The continuation is given
by $$N_{L,L'}^{-l-2} =-N_{L,L'}^l \quad{\rm and}\quad
N_{L,L'}^{-1}=N_{L,L'}^{k+1}=0\quad,$$ where $l=0,\cdots,k$. 
Thus the intersection number is given by the appropriate fusion
coefficient. Following ref. \cite{quintic}, we can write the
fusion coefficient $N_{L,L'}^M$ as a matrix in the index $M$. 
This matrix can be represented as a polynomial in $g$, the
generator of $\BZ_{k+2}$. Using this presentation of the fusion
coefficient, the Witten index for odd $k$ and $L=L'=(k-1)/2$ is
given by
\begin{equation}
{\cal I}^{mm} = \left( 1 + g + \cdots + g^{(k-1)/2} - g^{-1}-g^{-2} -\cdots
-g^{(-k-1)/2} \right)\quad,
\label{mmii}
\end{equation}
where we use the superscript $mm$ to denote that this is a
minimal model computation.

\subsection{Boundary Landau-Ginzburg Calculation}

We will now compute the Witten index in the LG model. The
worldsheet is assumed to have the topology of an annulus (of
width $\pi$).  We will impose A-type boundary conditions at the
two ends of the strip i.e., at $x^1=0$ and $x^1=\pi$.  We Wick
rotate the time coordinate to Euclidean space and make it
periodic.  At $x^1=\pi$. we impose the condition
\begin{eqnarray}
\partial_1{\rm Re}\ \phi &=& {\rm Im}\ \phi =0 \quad, \nonumber \\
\psi_+ &=& \overline{\psi}_-  \quad,
\end{eqnarray} 
This corresponds to the choice $A=1$ in the notation of the
earlier section.  At $x^1=0$, we impose
\begin{eqnarray}
\partial_1{\rm Re}\left(\exp(-{{i\pi m}\over{k+2}})\ \phi\right) &=& {\rm
Im}\left(\exp(-{{i\pi m}\over{k+2}})\  \phi\right) =0 \quad, \nonumber \\
\psi_+ &=& \exp({{2i\pi m}\over{k+2}})\ \overline{\psi}_-  \quad,
\end{eqnarray} 
This corresponds to the choice $A_m=\exp({{2i\pi m}\over{k+2}})$.

We will use the doubling trick to convert the annulus into a
torus. The doubling for fermions is done by introducing the
extended fermion $\Psi(x^1,x^0)$. 
\begin{equation}
\Psi(x^1,t) =\left\{\matrix{\psi_-(x^1,t)&{\rm for}&
0\leq\sigma\leq\pi\cr
\overline{\psi}_+(2\pi-x^1,t)&{\rm for} &
\pi\leq x^1 \leq 2\pi} \right.
\end{equation}
This automatically imposes the condition $\psi_- =
\overline{\psi}_+$ at the boundary at $x^1=\pi$. The boundary
condition at $x^1=0$ becomes the periodicity on the extended
fermion $\Psi$. The bosonic fields can also be doubled in a
similar fashion.

The counting of the Ramond ground states for the above situation
can now be seen to be identical to the counting of Ramond ground
states in the twisted sector of a certain orbifold: it is the
$m-$th twisted sector of the orbifolding of the $k-$th minimal
model by $Z_{k+2}$. This computation has been carried out by Vafa
and we quote his result\cite{vafa}.  For $m\neq0$, there is
precisely one ground state. This observation more or less
uniquely identifies the boundary condition with the $l= \lfloor
k/2\rfloor$ boundary states. From eqn. (\ref{mmii}), where we
have given the Witten index for the $l= \lfloor k/2\rfloor$
boundary states: one can clearly see that there is typically one
Ramond ground state in all sectors except in one case.  For
$m=0$, i.e., for the case where one has identical boundary
conditions on both ends of the annulus, one is dealing with the
untwisted sector. In this sector, free field methods cannot be
used. However, one has $(k+1)$ Ramond ground states. However,
none of them satisfy the $J_L=-J_R$ boundary condition in the
open-string channel. Thus, all Ramond ground states are projected
out and the Witten index is zero. (If $k$ were even, there is one
Ramond ground state with vanishing left and right $U(1)$ charges
and hence the Witten index is one in this case.)

In order to completely carry out the full LG computation, we need
to suitably assign a fermion number to the ground state. Let us
assign fermion number $(-)^m$ to the Ramond ground
state\footnote{The interchange of boundary conditions at
$x^1=0,\pi$ can be represented by $A_m\rightarrow A_{-m}$. This
choice makes the Witten index antisymmetric under the exchange.}
of the boundary condition given by $A_m$, for $m=1,\cdots,k+1$.
Using the conventions of Brunner et al., we can rewrite the above
result as (for odd $k$)
\begin{equation}
{\cal I}^{LG} =
\left(g + \cdots + g^{(k+1)/2} - g^{(k+3)/2}- \cdots -
g^{k+1}\right)\quad,
\end{equation}
where we use the superscript $LG$ to indicate that the Witten
index was computed in the LG model.  Note that
\begin{equation}
{\cal I}^{mm} = -g^{(k+1)/2} {\cal I}^{LG}\quad,
\end{equation} 
This difference can be understood as follows: The computation in
the LG model is a Witten index computation while the minimal
model computation is one where $\exp(i\pi J_0)$ replaces $(-)^F$.
Eqn.(\ref{relate}) provides the relation between the two
operations (on the boundary states).  Thus one interprets the
$\exp(i\pi J_0)$ to correspond to an additional time-twisting by
$g^{(k+3)/2}$ in the Witten index computation done in the doubled
theory.  Following the method used in closed string LG orbifolds
as in ref. \cite{vafa}, this can be seen to be be equivalent to
an additional space-twisting by $g^{-(k+3)/2}$ which provides the
required shift of $g^{(k+1)/2}$ in the calculation.  The minus
sign comes from the action of $h$, which maps branes to
anti-branes and thus changes the sign in the Witten index
computation. Thus, this identifies the LG boundary conditions
with the $L=(k-1)/2$ boundary states for odd $k$.

\subsection{Landau-Ginzburg Orbifolds}

The work of Greene, Vafa and Warner\cite{gvw} showed the
relationship between certain LG orbifolds and Gepner models. For
example, the Gepner model description of the Calabi-Yau
threefold, the quintic, at a special point in its moduli is given
by the tensor product of five copies of $k=3$ minimal models
subject to certain projections\cite{gepner}. The LG description
involves five chiral superfields $\Phi_i$ with superpotential
$$
W(\Phi) = \Phi_1^5 + \Phi_2^5 + \cdots + \Phi_5^5\quad.
$$
Further, the orbifolding corresponds to the identification
$\phi_i \sim \alpha \phi_i$, for all $i=1,\ldots,5$ ($\alpha$ is
a non-trivial fifth root of unity). It was argued by Greene, Vafa
and Warner\cite{gvw}, that this condition is equivalent to the
integer $U(1)$ projection in the corresponding Gepner model.

In the boundary LG, we find the following set of A-type boundary
conditions given by the matrix $A(\{m_i\})={\rm
Diag}(\alpha^{m_1},\ldots,\alpha^{m_5})$\cite{oldpaper}. The
equivalence relation mentioned earlier implies that the two
matrices given by parameters $\{m_i\}$ and $\{m_i +2\}$ are
equivalent. As before, we would like to calculate the Witten
index which is equivalent to the intersection matrix for these
cycles. 

Before orbifolding it is clear that the intersection matrix is
simply the product of the intersection matrix of the individual
theories. The further orbifolding by the diagonal $\BZ_5$(which
results in the integer U(1) charge projection) is implemented as
a projection (and hence time-twisting ) in the closed-string
channel. This will therefore show up as a sum over space-twisted
sectors in the open-string channel. Hence in the computation of
the Witten index in the doubled torus the final result is as
follows:
\begin{equation}
{\cal I}=\sum_{\nu=0}^{4} \prod_{i=1}^5 
N_{1,1}^{m_i-m'_i-4 + 2 \nu}
\end{equation}                                                  
where we $\{m_i\}$ and $\{m'_i\}$ are the boundary conditions
at the two ends of the annulus and $N_{1,1}^M$ is the $SU(2)_3$
fusion coefficent. The end result can be written
in the compact notation of ref. \cite{quintic} as
\begin{equation}
{\cal I}=\prod_{i=1}^5 (g_i + g_i^2 -g_i^3 -g_i^4)
\end{equation}
subject to the condition that $g_1g_2g_3g_4g_5=1$. The result is
as written in ref. \cite{quintic} and is consistent with the
$L=1$ assignment of boundary state labels for each individual
minimal model. 

The LG calculation, especially the part involving summing over
twisted sectors, is quite similar to the spacetime intersection
matrix calculation (see section 2 of ref. \cite{quintic}). In the
spacetime calculation, the intersection calculation involves
summing over patches which becomes different twisted sectors in
the LG orbifold computation.  For a single minimal model, we
needed to explain the shift between the $(-)^F$ and $\exp(i\pi
J_0)$ computations. However, for the LG orbifold, the two results
are identical due to the condition $g_1g_2g_3g_4g_5=1$.  Finally,
the similarity with the spacetime intersection calculation
suggests that the expectation from the modified geometric
hypothesis that the central charges of the A-branes should be the
same at different points in the K\"ahler moduli space is true. 

The methods used here clearly apply to more general examples
involving the linear boundary conditions in LG models as
discussed in section 2.  We will later see more general
conditions where the Witten index computation is quite difficult.

\section{Non-linear boundary conditions in LG models}

	As we have seen, the boundary LG description seems to
provide fewer boundary conditions than the corresponding boundary
minimal model.  This situation holds for more general examples
such as the one involving LG description of Calabi-Yau manifolds.
The class of boundary conditions considered in \cite{oldpaper}
correspond to the linear class.  We shall now try to generalise
these conditions and see if we can obtain new conditions.

\subsection{LG models with a single chiral superfield}

We shall first consider the simplest case of an LG model
involving a single chiral superfield $\Phi$. The most general
boundary condition is given by
\begin{equation}
F(\phi,\overline{\phi}) = 0\quad, 
\end{equation}
where $F(\phi,\overline{\phi})$ is a real function. We will have
to impose additional conditions such that A-type $N=2$
supersymmetry $\epsilon_+ = \eta \overline{\epsilon}_-$ is
preserved and all boundary terms (eqn. (\ref{ord})) which appear
in the ordinary variation of the Lagrangian vanish.  In order to
achieve this, we will first consider all new conditions generated
under the unbroken A-type $N=2$ supersymmetry.

The first supersymmetric variation leads to the following condition:
\begin{equation}
{{\partial F}\over{\partial\phi}}\ \psi_{+} + \eta\ {{\partial
F}\over{\partial\overline\phi}}\ 
\overline{\psi}_{-} =0 \quad.
\end{equation}
The supersymmetric variation of the above equation leads to the following
additional conditions: 
\begin{eqnarray}
\left[{{\partial F}\over{\partial\phi}}\ \partial_1 \phi -  {{\partial
F}\over{\partial\overline\phi}}\ 
 \partial_1 \overline{\phi}\right] - i 
\left[{{\partial F}\over{\partial\phi}}{{\partial F}
\over{\partial\overline\phi}}\right]^{1\over2} 
K \psi_{-}\overline{\psi}_- &=&0 \\ 
\left\{ F,W(\phi)-\overline{W}(\overline{\phi})\right\}_{PB} &=&0 \quad,
\end{eqnarray}
where $K$ is the extrinsic curvature of the curve $F=0$ in the complex
$\phi$-plane given by
$$
K = -
\left[{{\partial F}\over{\partial\phi}}{{\partial
F}\over{\partial\overline\phi}}\right]^{-{3\over2}}
\left [
\left({{\partial F}\over{\partial\overline{\phi}}}\right)^2 {{\partial^2F}\over {\partial\phi^2}}
-2 \left|{{\partial F}\over{\partial\phi}}\right|^2 {{\partial^2F}\over
{\partial\phi\partial\overline{\phi}}}
+\left({{\partial F}\over{\partial\phi}}\right)^2 {{\partial^2F}\over
{\partial\overline{\phi}^2}} \right]
$$
One can check that the boundary terms in the ordinary variation
vanish under these boundary conditions.  The linear cases
discussed in section 2 correspond to the case when $K=0$ (since
the boundary curves are straight lines in the $\phi$-plane) and
are clearly seen to be special case of the more general boundary
condition $F=0$. 

The vanishing of the Poisson bracket
$\{F,W(\phi)-\overline{W}(\overline{\phi})\}_{PB}$  imposes an important
restriction on the possible boundary curves in the $\phi$-plane.
Since in a two-dimensional phase space, there can be at most one constant
of motion, the only possible boundary condition is 
\begin{equation}
F=W(\phi)-\overline{W}(\overline{\phi}) -i c\quad,
\end{equation}
where $c$ is a real constant. These correspond to {\it straight
lines} in the W-plane which are parallel to the real $W$ axis.
For the case when $W=\phi^{k+2}/(k+2)$, the pre-image of $F=0$ in
the $\phi$-plane will generically have $(k+2)$ components. When
$c=0$, these $(k+2)$ pre-images are precisely the $(k+2)$ linear
boundary conditions that we have already obtained! 

We can now discuss as to how the other boundary conditions will
appear in the LG model. In this regard, we would like to make the
following observations: (i) The superpotential $W$ has $(k+1)$
degenerate minima at $\phi=0$. (ii) We will require that all the
curves $F=0$ should pass through the minima which fixes the
constant $c=0$. (iii) The minima can be made non-degenerate by
deforming the potential. A possible deformation is to add
$-\lambda \phi$ to the superpotential. This leads to
non-degenerate minima located at the $(k+1)$ roots of $\lambda$.
By a suitable rescaling, we can set $\lambda=1$. We will require
that the only allowed values of the constant $c$ are such that
$F=0$ passes through one of the minima.

Thus, we propose that the boundary states for $L=0,\cdots,k$
correspond to the boundary conditions in the LG model given by
the pre-images of the the straight lines in the W-plane:
$$
F_L(\phi,\overline{\phi})=W(\phi)-\overline{W}(\overline{\phi}) -i c_L =0\quad,
$$
where $c_L = 2 {\rm Im} W(\phi_L)$, where $\phi_L$ are the minima
of the bosonic potential. Each $F_L$ will have $(k+2)$ components
which will be asymptotic to the $k+2$ lines obtained in the
linear class of boundary conditions. This presumably should
enable us to associate them with the $M$ label of the boundary
states. Thus, the boundary states correspond to {\it real
algebraic curves} in the $\phi$ plane whose image in the $W$
plane are {\it straight lines} parallel to the Re$W$ axis. In the
degenerate case, it is easy to see that all $c_L$ are coincident.
Since we are as yet unable to compute the Witten index in these
non-linear situations, the identification cannot be made more
precise. 

\subsection{The general case}

We will now consider the general case of a LG model with $n$
chiral superfields and arbitrary superpotential. We will impose
$n$ independent conditions
\begin{equation}
F_a(\phi,\overline{\phi})=0\quad,
\end{equation}
where $F_a$ are real functions. We will use the indices
$i,j,\cdots$ to denote the superfields and the indices
$a,b,c,\cdots$ to indicate the boundary conditions.  Let $\Sigma$
denote the sub-manifold in $\BC^n$ (with complex coordinates
$\phi_i$ and $\overline{\phi}$) obtained by imposing these
conditions. We will in addition require that the functions be
compatible: 
\begin{equation}
\{F_a(\phi,\overline{\phi}),F_b(\phi,\overline{\phi})\}_{PB}=0\quad.
\end{equation}
We will assume that for all point on $\Sigma$, the normals
$\vec{n}_a \equiv (\partial_i F_a,\overline{\partial}_i F_a)$
span the normal bundle ${\cal N}\Sigma$. The vanishing of the
Poisson bracket can be rewritten as
\begin{equation}
\vec{n}_a \cdot \vec{t}_b =0
\end{equation}
where $\vec{t}_b\equiv (\partial_i F_a,-\overline{\partial}_i
F_a)$ are tangent vectors to the curve $F_b=0$. It follows that
they span the tangent bundle $T\Sigma$. Thus, $\Sigma$ is a {\it
Lagrangian submanifold} of $\BC^n$ by
construction\cite{harveylawson}. The induced metric (first
fundamental form) on $\Sigma$ is given by
\begin{equation}
h_{ab} = \vec{t}_a \cdot \vec{t}_b = \vec{n}_a \cdot \vec{n}_b \quad.
\end{equation}
Let $h^{ab}$ denote the inverse of the metric.

The first supersymmetric variation of the boundary conditions
leads to
\begin{equation}
{{\partial F_a}\over{\partial\phi_i}}\ \psi_{+i} + \eta\ {{\partial
F_a}\over{\partial\overline\phi_i}}\
\overline{\psi}_{-i} =0\quad, \label{fermbc}
\end{equation}
where the complex conjugate conditions are implicitly assumed. 
Defining
$$
\chi_{\pm a}\equiv {{\partial F_a}\over{\partial\phi_i}}\
\psi_{\pm i}\quad,
$$ 
the above condition takes the simple form
\begin{equation}
\chi_{+a} + \eta \overline{\chi}_{-a} =0\quad.
\end{equation}
Further supersymmetric variation of the above condition gives
rise to the following terms after imposing $\epsilon_+=\eta
\overline{\epsilon}_-$
\begin{eqnarray}
{{\partial^2 F_a}\over{\partial\phi_i\partial\phi_j}}
&&\left[ -\sqrt2 \epsilon_- \psi_{+j} + \sqrt2 \eta\overline{\epsilon}_-
\psi_{-j}\right]\psi_{+i} \nonumber \\
+{{\partial^2F_a}\over {\partial\phi_i\partial\overline{\phi}_j}}
&&\left[\sqrt2 \overline{\epsilon}_- \overline{\psi}_{+j} +
\sqrt2 \eta\epsilon_-\overline{\psi}_{-j} \right] \psi_{+i} \nonumber \\
+{{\partial F_a}\over{\partial\phi}_i}
&&\left[i\sqrt2(\partial_0 +\partial_1)\phi_i \overline{\epsilon}_- 
+ \sqrt2 \eta \overline{\epsilon}_- 
{{\partial \overline{W}}\over {\partial\overline{\phi}_i}}\right] 
\\
+\eta{{\partial^2F_a}\over {\partial\overline{\phi}_i\partial\overline{\phi}_j}}
&&\left[\sqrt2 \overline{\epsilon}_- \overline{\psi}_{+j} +
\sqrt2 \eta\epsilon_-\overline{\psi}_{-j} \right] \overline{\psi}_{-i} 
\nonumber \\
+\eta {{\partial^2F_a}\over {\partial\overline{\phi}_i\partial\phi}_j}
&&\left[ -\sqrt2 \epsilon_- \psi_{+j} + \sqrt2 \eta\overline{\epsilon}_-
\psi_{-j}\right] \overline{\psi}_{-i}  \nonumber \\
+\eta {{\partial F_a}\over{\partial\overline{\phi}_i}}
&&\left[i\sqrt2(\partial_0 -\partial_1)\overline{\phi}_i 
\eta\overline{\epsilon}_-
+ \sqrt{2} \overline{\epsilon}_- {{\partial W}\over
{\partial\phi_i}}\right]\nonumber =0\quad.
\end{eqnarray}
In the above, the terms involving $\partial_0\phi$ can be seen to
vanish using $\partial_0 F =0$. The remaining terms can be
rearranged in an elegant fashion using the extrinsic curvature
tensor $K_{abc}$ as defined in the appendix. After eliminating
$\chi_{+a}$ and $\overline{\chi}_{+a}$ respectively in terms of
$\overline{\chi}_{-a}$ and $\chi_{-a}$, we obtain
\begin{eqnarray}
\epsilon_- &&K_{abc}\  \chi_{-}^b\  \chi_{-}^c \nonumber \\
+  \overline{\epsilon}_- &&\left(
+i({{\partial F_a}\over{\partial\phi}_i}
\partial_1\phi_i
- {{\partial F_a}\over{\partial\overline{\phi}_i}}
\partial_1\overline{\phi}_i)
+K_{abc}\ \chi_{-}^b\ \overline{\chi}_{-}^c \right) \\ 
+\eta \overline{\epsilon}_-&& \left({{\partial F_a}\over{\partial\phi_i}}
{{\partial \overline{W}}\over {\partial\overline{\phi}_i}}
+ {{\partial F_a}\over{\partial\overline{\phi}_i}}
{{\partial W}\over {\partial\phi_i}}\right) =0\nonumber 
\end{eqnarray}
where $\chi_{-}^a = h^{ab} \chi_{-b}$.  In the above terms, it
can be seen that the term multiplying $\epsilon_-$ cancels since
a symmetric object multiplies an antisymmetric object. The terms
multiplying $\overline{\epsilon}_-$ lead to two new conditions
rather than a single one. There are two ways to understand this: 
First, the terms involving the superpotential are multiplied with
an $\eta$ and if we insist on a single condition, the bosonic
boundary condition ends up depending on the spin structure. In
addition, the vanishing of the boundary terms in the ordinary
variation also requires two conditions. The two conditions are
\begin{eqnarray}
\left(
\left[{{\partial F_a}\over{\partial\phi}_i}
\partial_1\phi_i
- {{\partial F_a}\over{\partial\overline{\phi}_i}}
\partial_1\overline{\phi}_i\right]
-i K_{abc}\ \chi_{-}^b\ \overline{\chi}_{-}^c \right) =0\\
\left\{F_a(\phi,\overline{\phi}),W(\Phi)-\overline{W}(\overline{\phi})
\right\}_{PB}=0
\end{eqnarray}
We note that the demonstration of the cancellation of the
boundary terms of the ordinary and supersymmetric variation of
the action is tedious but straightforward. The full set of
boundary conditions obtained by the the requirement of unbroken
$N=2$ supersymmetry of the A-type is equivalent to requiring that
the submanifold $\Sigma$ be {\it Lagrangian}. For the case
without a superpotential, this corresponds to the
microscopic(worldsheet)  realisation of situations considered by
Harvey and Lawson\cite{harveylawson}.  The {\it new feature} that
we obtain is that in the presence of a superpotential, there is
an additional condition that the real conditions $F_a$ have a
vanishing Poisson bracket with $(W-\overline{W})$. This suggests
that one must {\it necessarily} choose one of the conditions to
be $F=(W-\overline{W})-i c$ where $c$ is a real constant. This
can be seen as a consequence of the fact that in a phase space of
real dimension $2n$, there can only $n$ independent commuting
constants of motion.

\subsection{Spacetime supersymmetry and the special Lagrangian
condition}

The special Lagrangian condition\footnote{This was derived for
the first time using spacetime supersymmetry by ref. \cite{bbs}.
Of the subsequent literature on this approach, the one closest in
spirit to our microscopic viewpoint is ref. \cite{luest}.} which
is necessary for spacetime supersymmetric D-brane configuration
appears in our microscopic description as follows. (We will first
discuss the case when there is no superpotential. This is the
case where the spacetime is $\BC^n$). Let $\Upsilon$ and
$\overline{\Upsilon}$ respectively be the holomorphic $(n,0)$
form and anti-holomorphic $(0,n)$ form on $\BC^n$. In the
microscopic description, we can choose
\begin{eqnarray}
\Upsilon &\equiv& \psi_{-1}\psi_{-2} \cdots \psi_{-n}  \\
\overline{\Upsilon}&\equiv&
\overline{\psi}_{+1}\overline{\psi}_{+2} \cdots
\overline{\psi}_{+n}\quad,
\end{eqnarray} 
This choice is dictated by the fact that under the A-twist,
$\psi_{-i}$ become $(1,0)$ forms and $\overline{\psi}_{+i}$
become $(0,1)$ forms on $\BC^n$.  One can also see (by bosonising
the fermions, for example) that $\Upsilon$ generates spectral
flow in the left-moving $N=2$ under which the NS and R sectors
get mapped to each other. $\overline{\Upsilon}$ has a similar
action in the right moving sector except that it is a spectral
flow of opposite sign. The requirements for having boundary
conditions on the worldsheet which preserve spacetime
supersymmetry are:
\begin{itemize}
\item[(i)] The boundary
conditions must preserve a global $N=2$ worldsheet supersymmetry.
\item[(ii)] The boundary conditions must preserve a linear combination of
the two spectral flow generators\cite{ooy}. 
\end{itemize}
In our earlier considerations, we have ensured that the first part
has been satisfied. The second condition can be stated as follows:
\begin{equation}
\Upsilon = \eta^n\ \exp (i\alpha) \ \overline{\Upsilon}\quad,
\end{equation}
for some constant $\alpha$. Under the general A-type boundary
conditions discussed above, one can see that (using eqn.
(\ref{fermbc})) 
\begin{eqnarray}
\Delta\ \Upsilon &\equiv& \Delta\ \psi_{-1}\psi_{-2} \cdots \psi_{-n} 
\nonumber \\
&=& (-)^n \overline{\Delta} \
\overline{\psi}_{+1}\overline{\psi}_{+2} \cdots \overline{\psi}_{+n} \\
&=& (-)^n \overline{\Delta}\ \overline{\Upsilon} \quad, \nonumber
\end{eqnarray}
where $\Delta\equiv{\rm Det}\ {{\partial
F_a}\over{\partial\phi}_i}$.  The special Lagrangian condition
can now be restated as
\begin{equation}
\Delta = (-)^n \exp(i \alpha)\ {\overline{\Delta}} \quad.
\label{sL}
\end{equation}
The $(-)^n$ can always be absorbed into the phase $\alpha$ and
thus is not crucial. Hence the general boundary conditions which
preserve spacetime supersymmetry have to satisfy eqn. (\ref{sL}).

So far we have discussed special Lagrangian condition for the
case when the superpotential was zero, i.e., the target space was
$\BC^n$. In the presence of a superpotential, we have seen that
it is necessary to choose one of the conditions, say
$$
F_1= W(\phi)-\overline{W}(\overline{\phi})\quad.
$$ 
Further, let us assume that the superpotential is homogeneous and
that $\phi_i$ are homogeneous coordinates on
$\BP^{n-1}$\footnote{The generalisation to the case of
hypersurfaces in weighted projective spaces is obvious. We shall
restrict to projective spaces for simplicity.}. In order that the
boundary conditions carry over to $\BP^{n-1}$, we will require
that the $F_a$ be homogeneous under real scalings:
$\phi_i\rightarrow\lambda \phi_i$, for real $\lambda\neq0$.
Clearly, this is satisfied by $F_1=
W(\phi)-\overline{W}(\overline{\phi})$. 

Suppose, we have chosen $F_a$ which satisfy the conditions
mentioned in the previous paragraph and that the special
Lagrangian condition in $\BC^n$ given in eqn. (\ref{sL}) is also
satisfied. We will now show that this implies that one obtains a
special Lagrangian submanifold $\Sigma$ of the Calabi-Yau
manifold described by the equation $W=0$ in $\BP^{n-1}$.  The
global holomorphic $(n-2,0)$ form on the Calabi-Yau manifold is
given by\cite{candelas,atiyah}
\begin{equation}
\Omega = \int_\gamma {{\epsilon^{i_1\ldots i_n} \phi_{i_1}
d\phi_{i_2}\cdots d\phi_{i_n}} \over{W(\phi)}}\quad,
\label{holform}
\end{equation}
where $\gamma$ is a curve in $\BP^{n-1}$ enclosing $W=0$. On $\Sigma$,
the $F_a$ satisfy
\begin{equation}
dF_a =  
{{\partial F_a}\over{\partial\phi_i}}\ d\phi_{i} + {{\partial 
F_a}\over{\partial\overline\phi_i}} d\overline{\phi}_{i} =0
\end{equation}
Further, the homogeneity condition on the boundary conditions
$F_a$ can be written as
\begin{equation}
 \phi_{i}{{\partial F_a}\over{\partial\phi_i}} + 
\overline{\phi}_{i} 
{{\partial F_a}\over{\partial\overline\phi_i}} 
= d_a F_a\quad, \label{quasi}
\end{equation}
where $d_a$ is the degree of $F_a$.
Using the above two relations and the fact that we choose 
$W(\phi)=\overline{W}(\overline{\phi})$
as one of our boundary conditions, one can see that on $\Sigma$
\begin{equation}
\left. \phantom{{A\over B}}{\epsilon^{i_1\ldots i_n} \phi_{i_1}     
d\phi_{i_2}\cdots d\phi_{i_n}}\right|_\Sigma  =
\left. {{\overline{\Delta}}\over \Delta}
{\epsilon^{j_1\ldots j_n} \overline{\phi}_{j_1}                      
d\overline{\phi}_{j_2}\cdots d\overline{\phi}_{j_n}}\right|_\Sigma  
\end{equation}
This implies that
\begin{eqnarray}
\Omega|_{\Sigma} &=& (-)^n {{\overline{\Delta}}\over\Delta} \ 
\overline{\Omega}|_{\Sigma} \nonumber \\
&=& \exp(-i\alpha)\ \overline{\Omega}|_{\Sigma} 
\end{eqnarray}
which is the special Lagrangian condition on the Calabi-Yau
manifold.  Note that this has not been derived using any
spacetime inputs but rather from the worldsheet analysis of the
LG model with boundary. 
      
An important question to consider is whether the homogeneity
condition, eqn. (\ref{quasi}), which certainly appears natural,
is too restrictive. One possibility is to allow for the condition
\begin{equation}
 \left.\left(\phi_{i}{{\partial F_a}\over{\partial\phi_i}} + \exp(i\theta)
\overline{\phi}_{i} {{\partial F_a}\over{\partial\overline\phi_i}}\right)
\right|_\Sigma = 0\ {\rm mod}\ W\quad,
\end{equation}
where $\theta$ is a constant. The mod $W$ degree of freedom reflects the
fact that the integral in eqn. (\ref{holform}) has support only at the zeros
of $W$.  We will use this weaker condition shortly in an example. With
this weaker condition, one obtains
\begin{equation}
\Omega|_{\Sigma} 
= \exp(-i\alpha+i\theta)\ \overline{\Omega}|_{\Sigma} \quad.
\end{equation}

\subsection{Examples}

We first illustrate the case without a superpotential by using
the classic example of Harvey and Lawson\cite{harveylawson}. The
construction provides special Lagrangian submanifolds with
topology $\BR^+\times T^{n-1}$ on $\BC^n$. We will also show that
this leads naturally to a $T^{n-1}$ Lagrangian fibration of
$\BP^{n-1}$.

The conditions of Harvey and Lawson can be implemented as
boundary conditions in our worldsheet theory: 
\begin{eqnarray}
F_1&=&\left\{ \matrix{
{\rm Re} (\phi_1\ldots \phi_n)-c_1\quad{\rm for\ even}\ n \cr
{\rm Im} (\phi_1\ldots \phi_n)-c_1\quad{\rm for\ odd}\ n} \right. \\
F_a &=& |\phi_1|^2 - |\phi_a|^2 -c_a\quad {\rm for}\ 2\leq a \leq n
\end{eqnarray}
where $c_a$ are some real constants. Linear combinations of the
$c_a$ correspond to the radii of the circles of $T^{n-1}$.  When
$c_1=0$, the $\BR^+$ corresponds to the value of $|\phi_1|^2$ and
the $T^{n-1}$ corresponds to the $(n-1)$ phases left unfixed by
the condition $F_1=0$.

In order to extend these boundary conditions to $\BP^{n-1}$, the
condition of homogeneity on the $F_a$ implies that all the
constants $c_a$ must be necessarily set to zero.  In this limit,
the Lagrangian submanifold appears to be singular at the origin.
This however is not a point in $\BP^{n-1}$.  Thus, the
submanifold is non-singular. Further, the (real) scaling degree
of freedom in the homogeneous coordinates $\phi_i$ of $\BP^{n-1}$
eats up the $\BR^+$ degree of freedom leaving us with a
$T^{n-1}$. In the inhomogeneous coordinates of $\BP^{n-1}$, where
we set $\phi_1=1$, the radii of all circles is set to unity. Thus
the Lagrangian submanifold, $T^{n-1}$ in $\BP^{n-1}$ is obtained
at a specific point in its moduli space. We will momentarily see
how to generalise this. 

As pointed out by Strominger, Yau and Zaslow\cite{syz}, the
existence of a mirror partner for a Calabi-Yau threefold implies
that the Calabi-Yau manifold admits a $T^3$ fibration. Consider
the mirror quintic given by the equation in $\BP^4$: 
\begin{equation}
W = \phi_1^5 + \cdots + \phi_5^5 - 5 \psi \phi_1\phi_2\phi_3\phi_4\phi_5 
=0\quad.
\end{equation}
The large complex structure limit corresponds to
$|\psi|\rightarrow\infty$. In the infinite limit, the quintic
breaks up into five $\BP^3$ given by setting $\phi_i=0$ for
$i=1,\ldots,5$. While this is a degenerate limit, the $T^3$
special Lagrangian fibre is seen easily by using the earlier
construction for $\BP^3$. It has been argued by Strominger et.
al., that this $T^3$ will be special Lagrangian in the
neighbourhood of the infinite $|\psi|$ limit(see ref. \cite{ruan}
for a discussion). 

Since it in general rather hard to construct special Lagrangian
submanifolds, it is of interest to see how the above example fits
into our construction. In the infinite complex structure limit,
it is interesting to note that $F^1$ chosen by Harvey and Lawson
is indeed equal to $(W-\overline{W})$!  It immediately follows
therefore that if we also choose the conditions
\begin{equation}
F_a = |\phi_1|^2 - |\phi_a|^2 -c_a\quad {\rm for}\ 2\leq a \leq 5
\end{equation}
then we obtain a supersymmetric cycle. We have introduced four
constants $c_a$ which break the homogeneity condition.  However,
one can check that the weaker condition mentioned earlier holds: 
\begin{equation}
\left.\left(\phi_{i}{{\partial F_a}\over{\partial\phi_i}} -
\overline{\phi}_{i} {{\partial F_a}\over{\partial\overline\phi_i}}\right)
\right|_\Sigma = 0\ {\rm mod}\ W\quad,
\end{equation}
A calculation shows that except for $F_1=(W-\overline{W})$, the
above condition holds identically (without the mod $W$
condition).  It is of interest to count the number of independent
parameters.  First, it may seem that we have four angles and thus
a $T^4$. However, since $W=0$ necessarily requires one of the
$\phi_i$ to identically vanish, the angle associated with the
vanishing $\phi_i$ does not exist.  Further, projectivisation
leaves only three independent real variables coming from the four
$c_a$.  These presumably correspond to the moduli associated with
the $T^3$.  The example discussed above is not something specific
to the quintic but can be extended to a larger class of CY
threefolds. See for instance, ref. \cite{periods}.

\section{Conclusions and Outlook}

It is clear that the methods that we have outlined in this paper
have obvious generalizations. First, the methods can be extended
to LG models that are associated to CY hypersurfaces in weighted
projective spaces. Second, since we are working in the LG model
we can extend our techniques to hypersurfaces that are described
by more general potentials than those of the Fermat type.
Thirdly, the techniques could equally well be applied in the case
of non-quasi-homogeneous potentials relevant to massive $N=2$
supersymmetric theories.  There would however be some difference
in the geometric interpretation of the boundary conditions in
these cases. 

The special Lagrangian submanifolds considered here, described by
a set of real equations $F_a=0$ in some ambient $\BC^{n}$ can be
thought of as a real algebraic variety. This is in line with the
theory in the bulk corresponding to strings propagating on a
complex algebraic variety. It would be interesting to see how
other structures that were seen in the bulk, like the operator
ring for instance, carry over to the boundary theory. It is not
clear however what the potential role, if any at all, of
anholonomic boundary conditions, which of course are allowed in
general. While inequalities together with conditions of the form
$F_a=0$ are in general relevant to what are known to
mathematicians as semi-algebraic real sets. It would be
interesting to see if they have a role in the context of special
Lagrangian sub-manifolds in CY. 

One of the themes of this paper was to obtain boundary conditions
corresponding to all the boundary states in a single minimal
model. While the non-linear boundary conditions we have
constructed have suggested a possible way out, the story is far
from complete. It is of interest to be able to compute the Witten
index for the non-linear case in order to be sure of the
identification proposed in this paper.  Solving this problem is
of interest in making a clear geometric identification of the
Recknagel-Schomerus class of boundary states.  In particular, we
do not yet have a clear geometric identification of all the $L
\neq \lfloor k/2 \rfloor$ states in the boundary CFT.  We may add
that even in the linear class of boundary conditions we have not
yet explored the cases where the matrix A is non-diagonal. This
should help us in examining boundary states in the Gepner
construction that do not belong to the Recknagel-Schomerus class
of states. Such states would naturally arise from the possible
fixed-point resolutions of the modular transformation matrix of
the full Gepner CFT\cite{oldpaper}.

It is of interest to extend our analysis to the case of B-type
boundary conditions.  However, the LG description of B-branes
will be rather different from the large volume CY description
since the geometry and charge of the B-branes are not expected to
remain invariant. Nevertheless, it may be possible to track some
B-branes from the large volume CY limit to the LG phase without
encountering lines of marginal stability. Assuming that this is
possible, then one might be able to calculate the worldvolume
superpotential directly in the LG model.

As we discussed in section 2, it is not possible to impose
Neumann boundary conditions on all fields in the LG model. It is
also not possible to impose Neumann boundary conditions on the LG
field of a single minimal model.  This strongly suggests that all
the states of the Recknagel-Schomerus class must arise from
Dirichlet-type boundary conditions in the LG and suitable
modifications thereof. This shows that for example, a D6-brane
wrapping a CY will look rather different in the LG limit. From
the work of Brunner et. al. we know that the corresponding state
exists at the Gepner point in the moduli space. It would be of
interest to describe this state in the LG formalism. Given the
identification of the linear LG boundary conditions of A-type
with the $L=1$ boundary states of the A-type boundary CFT, the
case of B-type Dirichlet boundary conditions on all LG fields
described in section 2, most likely are the $\{L_i=1,\ {\rm for\
all}\ i \}$ B-type states. 

The B-type states may also be studied by using mirror symmetry on
A-type states in the LG model. In the LG models, the orbifolding
technique provides a simple method of constructing the mirror CY
and should also provide a corresponding method for the
construction of A-type boundary states in the mirror.

Given the close interplay of both complex and K\"ahler moduli in
the description of D-branes on CY threefolds (a natural
consequence of spacetime $N=2$ supersymmetry being broken to
$N=1$ by the D-branes), the linear sigma model(LSM) description
is better suited in some ways for a microscopic description on CY
threefolds. For example, one can show that the D6-brane (all
Neumann boundary conditions) in the CY limit starts looking like
an all Dirichlet boundary condition as one goes to ``small
volumes'' and to the LG phase. In the neighbourhood of vanishing
K\"ahler parameter (for the quintic), the CY as seen by the
D-brane appears to be in a non-commutative phase. These issues
will be discussed in a forthcoming paper. Related remarks appear
in the work of Hori and Vafa\cite{horivafa}.  The transitions
discussed by Joyce\cite{joyce} also seem well suited for an LSM
description. 

\noindent {\bf Acknowledgments} We would like to thank the
following for useful discussions: R. Balakrishnan,
V. Balakrishnan, G. Date,
S. Lakshmi Bala, D. S. Nagaraj, K. Paranjape and especially T. Sarkar.

\appendix
\section{Extrinsic curvature of Lagrangian submanifolds}

We will be considering a {\it Lagrangian submanifold} $\Sigma$ of
$\BC^n$ which is implicitly specified by $n$ independent {\it
real} functions
$$F_a(\phi,\overline{\phi})=0\quad,$$
where $\phi_i$ are complex coordinates on $\BC^n$.  The
Lagrangian condition implies that the Poisson bracket of the $n$
functions vanish\cite{harveylawson}. Further, the normals $n_a^i=
(\partial_i F_a, \overline{\partial}_i F_a)$ span the normal
bundle ${\cal N}\Sigma$ and the tangents $t_a^i= (\partial_i F_a,
-\overline{\partial}_i F_a)$ span the tangent bundle $T\Sigma$
and $T\BC^n = {\cal N}\Sigma\oplus T\Sigma$.  The vanishing
Poisson bracket ensures that $\vec{n}_a\cdot \vec{t}_b=0$. 

The tangential derivatives $D_a \equiv t_a^i\partial_i$ satisfy
$[D_a,D_b]=0$ by virtue of the vanishing of the Poisson bracket
$\{F_a,F_b\}_{PB}=0$. Thus, locally on $\Sigma$, there exists a
coordinate system $\sigma_a$ such that $\partial/\partial\sigma_a
= D_a$.  The induced metric({\it first fundamental form})  in
this coordinate system is given by
\begin{equation}
h_{ab} = \vec{t}_a \cdot \vec{t}_b = \vec{n}_a \cdot \vec{n}_b \quad.
\end{equation}
The extrinsic curvature tensor ({\it second fundamental form}) 
$\vec{K}_{ab}$ is defined as follows\footnote{We follow the
lectures of F. David\cite{david} in defining the extrinsic
curvature tensor after providing the required generalisations.}
\begin{equation}
(t_a^i\partial_i)\ t_b^j = (t_b^i\partial_i)\ t_a^j =
K_{ab}^j + \Gamma_{ab}^c\  t_c^j\quad,
\end{equation}
where $\Gamma_{ab}^c$ is the Christoffel connection with respect
to the induced metric on $\Sigma$.  Thus, $\vec{K}_{ab}$ is
normal to the surface $\Sigma$ (since the second term projects
out the tangential component of $(t_a^i\partial_i)t_b^j$). 
Since, $\vec{n}_a$ span the ${\cal N}\Sigma$, we can decompose
$\vec{K}_{ab}$ into
\begin{equation}
K_{abc}\equiv \vec{K}_{ab} \cdot \vec{n}_c\quad.
\end{equation}
One can explicitly calculate $K_{abc}$ as defined above and we obtain
\begin{equation}
K_{abc} =-\left [
{{\partial F_c}\over{\partial\phi_i}}
{{\partial F_b}\over{\partial\phi_j}}
{{\partial^2F_a}\over {\partial\phi_i\partial\phi_j}}
- {{\partial F_c}\over{\partial\overline{\phi}_i}} 
 {{\partial F_b}\over{\partial\phi_j}} 
{{\partial^2F_a}\over {\partial\phi_i\partial\overline{\phi}_j}}
- {{\partial F_c}\over{\partial\phi_j}} 
 {{\partial F_b}\over{\partial\overline{\phi}_i}} 
{{\partial^2F_a}\over {\partial\phi_i\partial\overline{\phi}_j}}
+{{\partial F_c}\over{\partial\overline{\phi}_i}}
{{\partial F_b}\over{\partial\overline{\phi}_j}}
{{\partial^2F_a}\over{\partial\overline{\phi}_i\partial\overline{\phi}_j}} \right]
\end{equation}
One can verify, that $K_{abc}$ is a completely symmetric tensor.
The symmetry under the exchange $a\leftrightarrow b$ is the usual
symmetry property of the extrinsic curvature tensor. However, for
{\it Lagrangian submanifolds}, one has the isomorphism between
the normal bundle and the tangent bundle which enables one to
make it fully symmetric\cite{bryant}.  Further, if $\Sigma$ is
special Lagrangian, then the trace of the extrinsic curvature
tensor with respect to the induced metric vanishes.

\end{document}